\documentclass[a4paper]{scrartcl}
\pdfoutput=1
\usepackage{array}
\usepackage{graphicx}
\usepackage{graphicx}
\usepackage{authblk}

\usepackage[english]{babel}
\usepackage[utf8]{inputenc}
\usepackage[tbtags]{amsmath}
\usepackage{amsfonts}
\usepackage{amssymb}
\usepackage{graphicx}
\usepackage[T1]{fontenc}
\usepackage{ae}
\usepackage[margin=10pt,font=small,labelfont=bf]{caption}
\usepackage{color}
\usepackage[plainpages=false]{hyperref}
\usepackage[numbers,sort&compress]{natbib}

\usepackage{microtype}
\usepackage{lmodern}

\usepackage{slashed}

\usepackage{dsfont}

\newcommand{\T}{\mathsf{T}}
\newcommand{\hc}{\mathrm{h.c.}}
\newcommand{\ii}{\mathrm{i}}
\newcommand{\ee}{\mathrm{e}}

\newcommand{\ex}[1]{\cdot 10^{#1}}

\newcommand{\unit}[1]{\,\text{#1}}

\newcommand{\eg}{~\textit{e.g.}}

\newcommand{\Equ}{Eq.}
\newcommand{\Equs}{Eqs.}
\newcommand{\Figu}{Fig.}
\newcommand{\Tab}{Tab.}

\newcommand{\reffig}[1]{\Figu~\ref{fig_#1}}
\newcommand{\refeq}[1]{\Equ~\eqref{eq_#1}}
\newcommand{\reftab}[1]{\Tab~\ref{tab_#1}}

\newcommand{\vectwo}[2]{\begin{pmatrix} #1 \\ #2 \end{pmatrix}}

\usepackage{booktabs}
\renewcommand{\toprule}{\specialrule{1.5pt}{0em}{0em}}
\renewcommand{\bottomrule}{\specialrule{1.5pt}{0em}{0em}}

\typearea{12}

\hypersetup{
        unicode = true,
        pdftitle = Higher Dimensional Effective Operators for Direct Dark Matter Detection , 
        pdfauthor = Martin Krauß and Stefano Morisi and Werner Porod and Walter Winter,
        colorlinks = true,
        linkcolor = blue,
        citecolor = blue,
        filecolor = blue,
        urlcolor = blue,
}

\newcommand{\AddrWue}{Institut f{\"u}r Theoretische Physik und Astrophysik, Universit{\"a}t W{\"u}rzburg,\\ Am Hubland, 97074 W{\"u}rzburg, Germany}

\title{\LARGE  Higher Dimensional Effective Operators \\ for \\ Direct Dark Matter Detection}

\author{\large Martin B. {Krauss}\footnote{\enspace Email: martin.krauss@physik.uni-wuerzburg.de}}

\author{\large Stefano {Morisi}\footnote{\enspace Email: stefano.morisi@physik.uni-wuerzburg.de}}

\author{\large Werner {Porod}\footnote{\enspace Email: porod@physik.uni-wuerzburg.de}}

\author{\large Walter {Winter}\footnote{\enspace Email: winter@physik.uni-wuerzburg.de}}

\affil{\small \AddrWue}

\date{\large February 14, 2014}

\begin{document}

\maketitle

\begin{abstract}
\begin{center}
{\textbf{Abstract}}
\end{center}
We discuss higher dimensional effective operators describing interactions between fermionic dark matter and Standard Model particles. They are typically suppressed compared to the leading order effective operators, which can explain why no conclusive direct dark matter detection has been made so far.
The ultraviolet completions of the effective operators, which we systematically study, require new particles.
These particles can  potentially have masses at the TeV scale and can therefore be phenomenologically interesting for LHC physics.
We demonstrate that the lowest order options require Higgs-portal interactions generated by dimension six operators. We list all possible tree-level completions with extra fermions and scalars,
and we discuss the LHC phenomenology of a specific example with extra heavy fermion doublets.
\end{abstract}

\newpage

\section{Introduction}
Direct searches for dark matter particles have been so far inconclusive. 
Hints for a possible observation of a DM particle with a mass around 6 to 10 GeV have been found by CoGeNT~\cite{Aalseth:2010vx} and CDMS~\cite{Agnese:2013rvf,Ahmed:2010wy}. After the recent revision of the XENON10 results~\cite{Frandsen:2013cna,Angle:2011th_err}, the constraints on the CoGeNT/CDMS signal region from XENON10 have been significantly weakened. There is, however, a tension with regard to the constraints from XENON100. So far it is not clear how to resolve this conflict, but possible models to avoid it have been discussed in the literature~\cite{Frandsen:2013cna,Mao:2013nda,Cotta:2013jna,Baek:2012se}. One possibility is to have isospin violating dark matter~\cite{Kurylov:2003ra,Giuliani:2005my,Chang:2010yk,Kang:2010mh,Feng:2011vu,Feng:2013vod} that can suppress interactions with Xenon. It has also been suggested, that uncertainties about the sensitivity of the XENON100 experiment may relax this tension~\cite{Hooper:2013cwa}. To learn more about the true nature of DM and whether it exists in the indicated parameter region, more observational data are required.  Recently the LUX~experiment~\cite{Akerib:2012ys} has published new results~\cite{Akerib:2013tjd} giving further constraints. The strongest bounds from collider experiments are mono-jet and mono-photon searches at the LHC~\cite{ATLAS:2012ky,Chatrchyan:2012me,Aad:2012fw,Chatrchyan:2012tea,Zhou:2013fla}. These searches are sensitive for the pair production of DM particles with initial state radiation. For a recent review on dark matter searches 
see,\eg, Ref.~\cite{Bauer:2013ihz}.

As the fundamental theory behind dark matter (DM) is not yet known, it is a frequently used approach to describe interactions of DM particles by using effective operators~\cite{Bai:2010hh,Fox:2011fx,Beltran:2008xg,Agrawal:2010fh,Zheng:2010js,Yu:2011by,Goodman:2010yf,Goodman:2010ku,Buckley:2013jwa}. A better understanding of the effective operators describing DM interactions will help in the discussion of those experiments and their results. In particular, the fundamental theory leading to such effective operators gives rise to alternative ways to test the model, and, in fact, there may be observational implications showing up elsewhere. It is therefore interesting to study the high energy completions of the effective theory in a systematic way; see,\eg, Ref.~\cite{Dreiner:2012xm,Dreiner:2013vla,Schmeier:2013kda} for the lowest order decompositions.

One possible explanation for not having detected dark matter yet is that the DM interaction may be governed by a higher dimensional operator. Higher dimensional operators are an effective parametrization of physics at a heavy scale. They are suppressed by powers of the new physics scale $\Lambda$:
\begin{align}
 \mathcal{L} = \mathcal{L}_{\rm SM} + \mathcal{L}^{d=5}_{\text{eff}} 
+ \mathcal{L}^{d=6}_{\text{eff}} + \cdots
\, , \quad \textrm{with} \quad \mathcal{L}^{d}_{\text{eff}} \propto \frac{1}{\Lambda_{\mathrm{NP}}^{d-4}} \, \mathcal{O}^{d}\,.
\end{align}
In the case of DM interactions, the additional suppression of the higher dimensional operators can reduce the cross-section of detection processes even if new physics appears at the TeV scale and the couplings are order unity. In that case, the correct relic abundance, as observed by WMAP and PLANCK~\cite{Komatsu:2010fb,Ade:2013zuv}, can still be obtained. The usual mechanism to obtain these abundances is thermally produced DM. In this case DM particles are in thermal equilibrium until their annihilation rate becomes insufficient and they freeze out. An alternative model is that of a feebly interacting massive particle (FIMP) that is produced via a so-called freeze-in mechanism~\cite{Hall:2009bx}: The particle has initially a low abundance. Due to its very weak interactions with the particles in the thermal bath it cannot reach the equilibrium state. Its abundance will increase, nonetheless, due to these interactions, until the temperature drops sufficiently below its mass.

To generate higher dimensional operators, additional mediators and interactions will be needed, which means that additional phenomenology is implied. We therefore focus on the possibility to have higher dimensional operators as leading contribution to DM interactions, and we systematically discuss the possible implementations of such operators in the following at tree-level. We will show that many of these models can be ruled out or be constrained on the basis of very general and model-independent arguments. 

We focus on fermionic dark matter in this study, and we consider higher dimensional operators with extra SM singlet scalars and fermions. This study is organized as follows: We recapitulate the decompositions of the leading order effective operators in Sec.~\ref{sec_effOps}, which we need later to exclude mediators or interactions present at leading order. Then we discuss direct detection from higher dimensional operators and their tree-level decompositions in Sec.~\ref{sec:higher}. Finally, we illustrate possible tests at the LHC for one example with extra fermion doublets in Sec~\ref{sec:lhc}.  

\section{Leading order effective operators}
\label{sec_effOps}

\begin{figure}[t]
 \begin{center}
\vspace*{1cm}
  \begin{tabular}{ccc}
    \includegraphics[scale=1.2]{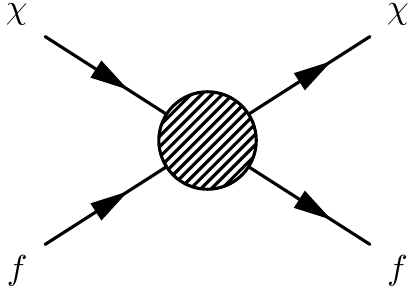}%
& \hspace*{2cm} &
    \includegraphics[scale=1.2]{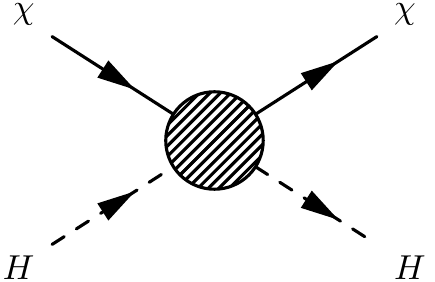}%
\\
 $\mathcal{O}_1$ & \ & $\mathcal{O}_H$
  \end{tabular}
 \end{center}
\caption{Lowest order dark matter interactions considered in this study.} \label{figintr}
\end{figure}

\begin{figure}[t]
 \begin{center}
\vspace*{1cm}
  \begin{tabular}{ccc}
   \begin{minipage}{.35\linewidth}
    \includegraphics[scale=1.2]{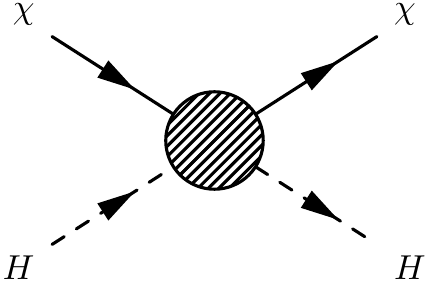}%
\end{minipage}
& $\qquad\longrightarrow\qquad$ &
   \begin{minipage}{.35\linewidth}
    \includegraphics[scale=1.2]{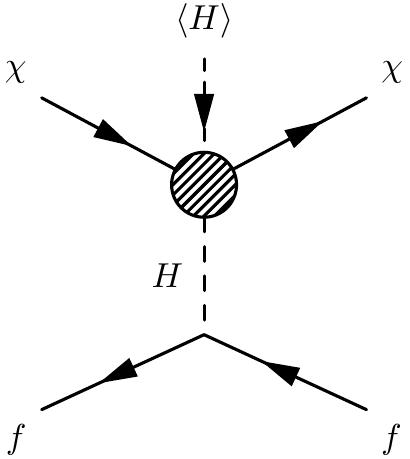}%
\\
\end{minipage}
\\
$\mathcal{O}_H$ & \ & $\mathcal{O}_2$
  \end{tabular}
 \end{center}
\caption{Leading order interactions of fermionic DM with SM particles: Direct detection interaction $\mathcal{O}_2 = 1/(\Lambda m_H^2) \, \chi\chi ff \langle H \rangle$ (r.h.s) generated from the  Higgs-portal  effective operator $\mathcal{O}_H = 1/\Lambda \, \chi\chi  H^\dagger H $ (l.h.s) from the SM Yukawa interaction.} \label{fig_HPortal}
\end{figure}

The following classes of operators describe all leading order interactions of DM fermions, denoted as $\chi$, with SM particles, see \Figu~\ref{figintr}:
\begin{align}
\mathcal{O}_1 = &\frac{1}{\Lambda^2} \chi\chi \,f f\label{eq_o1}\,, \\
\mathcal{O}_H =  &\frac{1}{\Lambda} \chi\chi  H^\dagger H \label{eq_oh}\, .
\end{align}
In literature these operators are discussed to describe interactions of DM with SM fermions. Operator $\mathcal{O}_1$ is a dimension six operator describing the interactions among DM fermions and SM fermions directly. Operator $\mathcal{O}_H$ is a dimension five operator describing interactions among DM fermions and the SM Higgs. This so-called Higgs-portal operator
\begin{equation}
\frac{\lambda_{hff}}{\Lambda} \chi\chi \,H^\dagger H\label{e3} \,,
\end{equation}
is well known and studied in the literature~\cite{Patt:2006fw, Kim:2006af,
MarchRussell:2008yu, Kim:2008pp, Ahlers:2008qc, Feng:2008mu, Andreas:2008xy,
Barger:2008jx, Kadastik:2009ca, Kanemura:2010sh, Piazza:2010ye,
Arina:2010an, Low:2011kp, Djouadi:2011aa, Englert:2011yb, Kamenik:2012hn,
Gonderinger:2012rd, Lebedev:2012zw,Baek:2011aa,Baek:2012se}. Recent detailed analysis \cite{Djouadi:2011aa, Djouadi:2012zc, LopezHonorez:2012kv} 
have found constraints for the effective coupling $\lambda_{hff}$ mostly from XENON100, namely
\begin{equation}
\frac{\lambda_{hff}}{\Lambda} \lesssim 1\ex{-3}\unit{GeV}^{-1}\,.\label{lambdahff}
\end{equation}
 As we illustrate in \Figu~\ref{fig_HPortal}, the SM Yukawa interactions generate the following direct detection $d=6$ operator after electroweak symmetry breaking:
\begin{equation}
\mathcal{O}_2 =  \frac{1}{\Lambda m_H^2} \, \chi\chi ff \langle H \rangle \label{eq_o2} \, .
\end{equation}
In spite of the higher dimension of that operator than \refeq{o1}, the two operators can be of the same order, and, in fact, \refeq{o2} may be even leading. The reason is that the SM Higgs takes the role of one of the mediators on the r.h.s. of \Figu~\ref{fig_HPortal}. In the following we will discuss this in more detail: The effective coupling of DM particles to nucleons, which is experimentally relevant, is for $\mathcal{O}_1$ given by
\begin{align}
 \lambda^\text{N,eff}_{\mathcal{O}_1} \chi \chi ff\,,
 \quad \text{with} \quad
\lambda^\text{N,eff}_{\mathcal{O}_1} = f_N^{\mathcal{O}_1} \frac{1}{\Lambda^2}
\end{align}
with the form-factor $f_N^{\mathcal{O}_1}$ and assuming order one couplings. For $\mathcal{O}_2$ we obtain
\begin{align}
\lambda^\text{N,eff}_{\mathcal{O}_2} \chi \chi ff\,,
 \quad \text{with} \quad
\lambda^\text{N,eff}_{\mathcal{O}_2} = f_N^{H} \frac{\langle H \rangle}{\Lambda m_H^2}
\end{align}
where all non-SM couplings are assumed order one and flavor blind. The nuclear form-factor for the coupling of the Higgs to nucleons is given by~\cite{Kanemura:2010sh}
\begin{align}
 f_N^H = \sum_q f_{Tq} + \frac{2}{9} f_{TG}\,,
\end{align}
where $m_N$ is the nucleon mass, $f_{Tq}$ are the form factors for the coupling of the Higgs to the quarks in the nucleon, known from Lattice-QCD and $f_{TG}$ is the gluon form-factor. Numerically $f_N^H$ is of about order one.
The form factor for the direct interaction can be adopted from the one used for effective spin-independent neutralino-nucleon cross-sections (see,\eg, Ref.~\cite{Falk:1999mq}), which correspond to operator $\mathcal{O}_1$:
\begin{align}
 f_N^{\mathcal{O}_1} = m_N \left(\sum_{q=u,d,s} \frac{f_{Tq}}{m_q} + \frac{2}{27} f_{TG} \sum_{q=c,b,t} \frac{1}{m_q}\right)\,.
\end{align}
Its numerical value is also order one. In conclusion, we find that the direct DM-nucleon interaction via $\mathcal{O}_1$ is roughly suppressed by a factor $m_H / \Lambda$ as compared to the Higgs-portal induced interaction of $\mathcal{O}_2$.

We therefore observe that any higher dimensional operator as natural extension of the operator $\mathcal{O}_1$, must not imply the existence of $\mathcal{O}_1$ itself \textit{or} $\mathcal{O}_H$ in order to be considered the dominant contribution to DM interactions. There are three generic possibilities to circumvent this issue: (i) either a symmetry (a so-called ``matter parity'') is used to forbid the lowest order operators, or (ii) the mediators or interactions are chosen such that none of the operators in \Figu~\ref{figintr} will appear, or (iii) a combination of (ii) with a symmetry to forbid certain interactions is implemented. Below we will demonstrate that option (i) does not generically work, since some of the Lorentz configurations are invariant under any such symmetry. Option (ii), on the other hand, requires that the decompositions of the operators in \Figu~\ref{figintr} be known. Therefore, we list all possibilities in terms of the SM quantum numbers of the mediators below. We use the notation $X^{s/f}_{Y}$ for the mediators where $X$ is the SU(2) nature, the superscript 
denotes scalar (s) or fermion (f) and $Y\equiv Q - I_3^W$ is the hypercharge.

In general, there exist several possible renormalizable theories which can lead to the same effective operator. In such a  fundamental theory new heavy mediator fields are introduced. If they are integrated out, the corresponding effective operator will be generated. For the decomposition of the effective operators, we use the techniques which have been applied to neutrino masses \cite{Bonnet:2009ej, Krauss:2011ur}, neutrinoless double beta decay \cite{Bonnet:2012kh}, and anomalous Higgs couplings \cite{Bonnet:2011yx} before. We discuss the Lorentz structures and decompositions for \refeq{o1} and \refeq{oh} in the next following subsections.

\subsection{Effective operators of the class $\chi \chi \,ff$}

In this case one has to distinguish between Dirac and Majorana dark matter. In the first case, the dark matter is composed of two (Weyl spinor) components $X = (\chi_R$,$\chi_L)$, which have the mass term
\begin{align}
m_\chi^\text{Dirac} \, \overline{X_R} X_L =  m_\chi^\text{Dirac} \chi_R^c \chi_L\,. \label{eq_mDiracDM}
\end{align}
In the second case, we only have one Weyl component $X = (\chi^c, \chi)$. The mass term is
\begin{align}
 m_\chi^\text{Maj} \overline{X_L^c} X_L =  m_\chi^\text{Maj} \chi \chi \,, \label{eq_mMajDM}
\end{align}
where $X_{R/L} = (1 \pm \gamma_5) X$. We use the convention that the product of left-handed Weyl spinors is understood to be Lorentz invariant, thus $\chi \chi \equiv \chi^\T \ii \sigma_2 \chi$ and $\chi_R^c \chi_L \equiv (\chi_R^c)^\T \ii \sigma_2 \chi_L$ is the invariant product of two left-handed Weyl-spinors.

The lowest dimensional operators describing the scalar and vector interactions of dark matter with SM fermions can be related by Fierz identities and are
\begin{subequations}
\begin{align}
(\overline{X_L} f_R) \ (\overline{X_R} (f_R)^c) &= \frac{1}{2}(\overline{X_L} \gamma^\mu (f_R)^c)(\overline{X_R} \gamma_\mu f_R)\\
\overline{X_L} (f_L)^c \ \overline{X_R} f_L     &= \frac{1}{2}(\overline{X_L} \gamma^\mu f_L)(\overline{X_R} \gamma_\mu (f_L)^c) \\
\overline{X_L} f_R \ \overline{f_R} X_L         &= \frac{1}{2}(\overline{X_L} \gamma^\mu X_L) (\overline{f_R} \gamma_\mu f_R)\\
\overline{X_R} f_L \ \overline{f_L} X_R         &= \frac{1}{2}(\overline{X_R} \gamma^\mu X_R) (\overline{f_L} \gamma_\mu f_L) \\
(\overline{X_L} \gamma^\mu f_L)(\overline{f_L} \gamma_\mu X_L) &= -(\overline{X_L} \gamma^\mu X_L)(\overline{f_L} \gamma_\mu f_L)\\
(\overline{X_R} \gamma^\mu f_R)(\overline{f_R} \gamma_\mu X_R)&= -(\overline{X_R} \gamma^\mu X_R)(\overline{f_R} \gamma_\mu f_R)
\end{align} \label{eq_OpDiracDM}
\end{subequations}
The factors $1/\Lambda^2$ are understood to be present everywhere. 
In \refeq{OpDiracDM}, terms such as $\overline{X_R} X_L \overline{f_R} f_L$ are forbidden by the SM gauge group. One can easily see that \mbox{operators~(\ref{eq_OpDiracDM}c-f)} were even invariant if additional Abelian symmetries were present, since all fields appear as conjugated pairs. In the Majorana case, operators (\ref{eq_OpDiracDM}a-f) are invariant under symmetries. This is due to the fact that left- and right-handed components are not independent. Thus, in summary, it is not possible to forbid all of these operators by a symmetry. We will discuss only Dirac DM in the following, but the discussion of Majorana DM can be made in a parallel way by replacing $\chi_R$ with $(\chi_{L})^c$. In the following we will mostly use Weyl spinors.

\begin{figure}[t]
\begin{center}
\begin{tabular}{ccc}
    \includegraphics[scale=1.2]{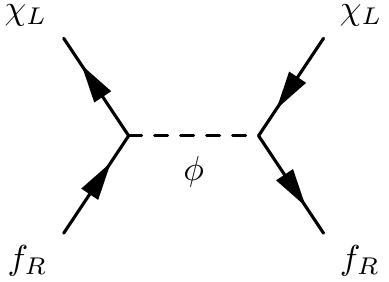}%
& \hspace*{5mm}
    \includegraphics[scale=1.2]{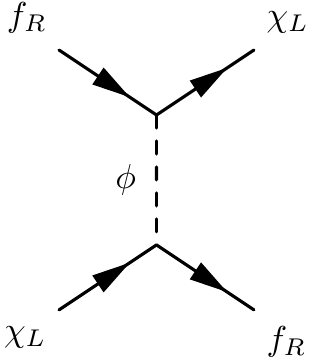}%
&
    \includegraphics[scale=1.2]{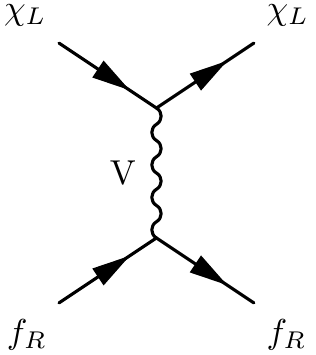}%
\\ \ \\
\#A1 & \#A2 & \#A3
\end{tabular}
\end{center}
 \caption{Decompositions of Operator $(\overline{X_L} f_R) \ (\overline{f_R} X_L) = \frac{1}{2}(\overline{X_L} \gamma^\mu X_L) (\overline{f_R} \gamma_\mu f_R)$ from \refeq{OpDiracDM}}
\label{fig_decompD6}
\end{figure}

In order to check if the mediators and interactions needed to construct a higher dimensional operator will lead to any of the two operators in \Figu~\ref{figintr}, we exemplify the decompositions in \reffig{decompD6} for operator $(\overline{X_L} f_R) \ (\overline{f_R} X_L)$; it is similar for the other ones  (see also \Tab~I of Ref.~\cite{Dreiner:2012xm}). The decomposition with vector bosons can be easily avoided if $\chi$ is a singlet under any gauge symmetry. The scalar mediator has to carry exactly the same (or conjugated) quantum numbers as the external SM fermion $f_R$ under the SM gauge symmetry, since $\chi$ is a SM singlet. In order to stabilize the DM we assume it is charged under a $\mathbb{Z}_2$ symmetry whereas all SM particles are not. We will use the notation ``+'' for $\mathbb{Z}_2$-even particles and ``-'' for $\mathbb{Z}_2$-odd particles. From the diagrams in \reffig{decompD6} one can easily see that also the scalar mediators will be odd under this parity. In conclusion we find that the operators from \refeq{OpDiracDM} are only present in models that include additional vector bosons and scalars with the following charges under $(SU(3)_c,SU(2)_L,U(1)_Y;\mathbb{Z}_2)$:
\begin{subequations}\label{eq_qMediators}
 \begin{align}
  &\ \left(3,*,*;-\right)\,, \\
  &\left(1,2,-\frac{1}{2};-\right)\,,\\
  &\ \left(1,1,-1;-\right) \, ,
 \end{align}
\end{subequations}
or their charge conjugates; the $*$ refers to any possible charge. We will show in 
section~\ref{sec:lhc} a realization of such a model.

\subsection{Effective operator $\chi \chi \,H^\dagger H$}

\begin{figure}[tb]
\begin{center}
\begin{tabular}{cc}
    \includegraphics[scale=1.2]{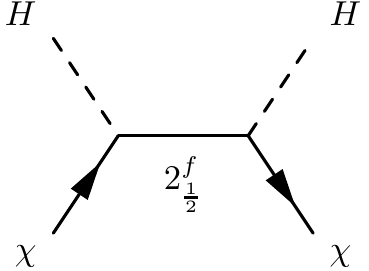}%
& \hspace*{5mm}
    \includegraphics[scale=1.2]{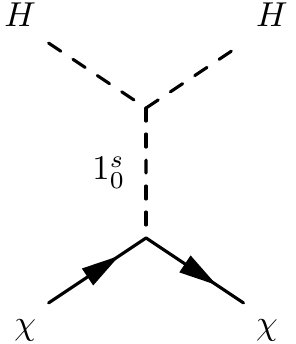}%
\\ \ \\
\#H1 & \#H2
\end{tabular}
\end{center}
 \caption{Decompositions of the effective operator $\chi \chi \, H^\dagger H$.}
 \label{fig_HPdecomp}
\end{figure}

The effective operator $\mathcal{O}_H$ is even simpler to decompose, because there is only one possible Lorentz nature. The decompositions  are shown in \reffig{HPdecomp}. The mediators are (in the notation of \refeq{qMediators})
\begin{subequations}
\begin{align}
&\left(1,2,\pm\frac{1}{2},-\right) \qquad \text{and}\\
&\ \left(1,1,0;+\right)\,.
\end{align}
\end{subequations}

\section{Higher order effective operators, and their decompositions at tree level}
\label{sec:higher}

\begin{figure}[t]
 \begin{center}
\vspace*{1cm}
  \begin{tabular}{ccc}
   \begin{minipage}{.25\linewidth}
    \includegraphics[scale=1.2]{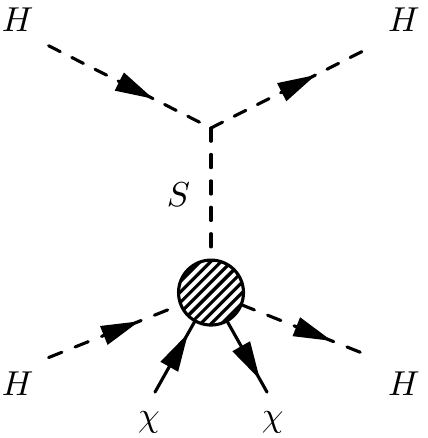}%
\end{minipage}
& $\qquad\longrightarrow\qquad$ &
   \begin{minipage}{.25\linewidth}
    \includegraphics[scale=1.2]{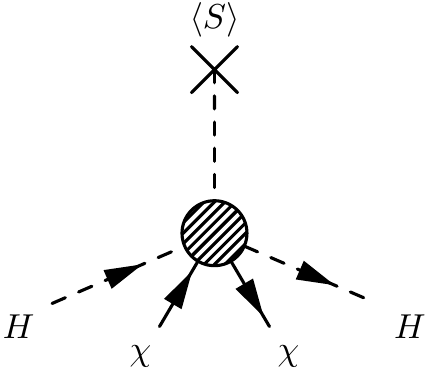}%
\end{minipage}
  \end{tabular}
 \end{center}
\caption{Schematic illustration how decompositions of the type $\chi\chi( H^\dagger H)^2$ with a scalar singlet mediator $S$ will induce operators of the type $\chi\chi H^\dagger H S$ if $S$ obtains a VEV.} \label{fig_HHvS}
\end{figure}

In \reftab{ops01} we list all operators up to d=7 that are relevant for DM interactions, in the direct interaction as well as in the Higgs-portal scenario. Here we include also operators with a scalar singlet $S$ as external field, although such a field is not part of the SM. As will be seen in the following, $S$ appears, however, in many decompositions of operators of the type $\chi\chi (H^\dagger H)^2$ as a mediator that couples to the Higgs field via a term $H^\dagger H S$. Since the $S$ can obtain a VEV due to this coupling, an operator of the type $\chi\chi H^\dagger H S$ will be induced. This is schematically illustrated in \reffig{HHvS}. One therefore should also consider the latter operators in order to present a consistent picture of higher dimensional DM interactions.

In the following we will discuss the next-to-leading order operators for direct interactions, the $d=7$ operators $\chi \chi f f S$ and $\chi \chi f f H$. We will show that they always imply that also the leading direct or Higgs-portal operators exist and that they therefore give only sub-dominant contributions to DM interactions. We will therefore focus on the next-to-leading Higgs-portal operator $\chi \chi H^\dagger H S$. We will also discuss the operator $\chi \chi (H^\dagger H)^2$, which is the next-to-leading order operator that does only contain SM fields.

\begin{table}[tb]
\begin{center}
 \begin{tabular}{lcc}
\toprule
		& (a) 			& (b)   \\
\midrule
$d=5$ 		& ---			& $\chi \chi H^\dagger H$  	\\
\midrule
$d=6$		& $\chi \chi f f$	& $\chi \chi H^\dagger H S$	\\
\midrule
$d=7$		& $\chi \chi f f S$	& $\chi \chi (H^\dagger H)^2$	\\
		& $\chi \chi f f H$	& $\chi \chi H^\dagger H S^2$	\\
\bottomrule

 \end{tabular}
\end{center}
\caption{Higher dimensional operators generating dark matter interactions (a) by direct interactions and (b) via the Higgs-portal.}
\label{tab_ops01}
\end{table}

While the suppression of the interaction of DM with SM particles via higher dimensional operators is favored by experimental bounds, small cross-sections make it difficult to obtain the observed abundances for thermally produced DM. A small Higgs-portal interaction can still be in agreement with cosmological bounds if the DM has masses close to half of the Higgs mass leading to resonant production~\cite{LopezHonorez:2012kv}. In that case one has also to consider limits from the Higgs to invisible branching ratio~\cite{Djouadi:2011aa}. Another possibility is that a lower dimensional operator is responsible for production but is not involved in direct detection, since it does not couple to quarks. We will discuss a possible realization of such a mechanism in Sec.~\ref{sec:cchhs}.
Finally, it is also possible that DM is produced non-thermally. In this context it is interesting to discuss feebly-interacting massive particles (FIMPs). In principle, higher dimensional operators and their additional suppression of interactions can be a way to generate small interactions as required for FIMPS. This works only if the mediators are not present anymore in the thermal bath, i.e., they have to be very heavy. If they were present, the FIMP particles would couple to the mediators in the bath, which would bring them into thermal equilibrium and destroy the freeze-in mechanism. For mediators at the  TeV scale, this would be a problem.

\subsection{Operators of the type $\chi \chi \,ff S$}

\begin{figure}[t]
\begin{center}
\begin{tabular}{ccc}
    \includegraphics[scale=1.2]{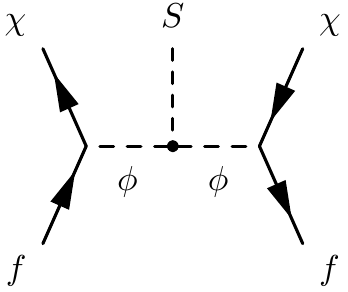}%
&
    \includegraphics[scale=1.2]{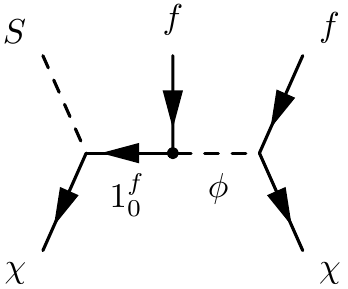}%
&
    \includegraphics[scale=1.2]{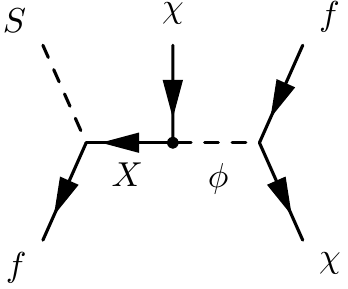}%
\\ \ \\
\#S1 & \#S2 & \#S3
\end{tabular}
\end{center}
\caption{Different decompositions (Feynman diagrams) for the operator $\chi \chi ff S$. $\phi$ is a scalar and $X$ a fermion that has the quantum numbers of the external fermions $f$. $\phi$ corresponds to the mediators in \reffig{decompD6}.}
\label{fig_decompS}
\end{figure}

The decompositions of the operator $\chi \chi \,ff S$ can be easily obtained from operator \#A1 (or \#A2) in \reffig{decompD6} by inserting a coupling to the external $S$ field at all possible lines. Note that  we assume the DM field $\chi$ to be a gauge singlet, which means that we do not have to consider direct couplings to vector bosons at tree-level, such as in \#A3. The result is shown in \reffig{decompS}. Not all decompositions that would be possible for generic fermion fields are possible, because SM fields ought to conserve SU(2) and chirality at the same time. One can easily see that all fields and couplings that generate the $d=6$ operator $\chi \chi f f$ are also present in any decomposition of the $\chi \chi \,ff S$ operator, so that the existence of this operator would also imply the existence of the leading order direct interaction operator. This can be demonstrated by looking at the Lagrangian of operator \#S1:
\begin{align}
\mathcal{L}_\text{\#S1} = \mathcal{L}_\text{SM}
+ \lambda_{\chi f \phi}\, \chi f \cdot \phi
+ \lambda_{S\phi\phi}\, S^\dagger \phi \cdot \phi
+ m_\phi\, \phi^\dagger \phi
+ m_\chi\, \chi \chi
+ \dotsb
+ \hc\,,
\end{align}
where the dot between two fields signifies a product that is invariant under the SM gauge group, which depends on the quantum numbers of $f$. One can now see that the terms $\lambda_{\chi f S} \chi f \cdot \phi$ and $m_\phi \phi^\dagger \phi$ will also lead to the diagrams $\#A1$ and $\#A2$ of \reffig{decompD6}. A similar argument can be made for operators $\#S1$ and $\#S2$. We conclude that any of the decompositions $\#S1$, $\#S2$ or $\#S2$ would also imply the existence of operator $\#A1$ or $\#A2$. Or in other words, if the operator $\chi \chi \,ff$ is forbidden, also the operator $\chi \chi \,ff S$ will not be present.

\subsection{Operators of the type $\chi \chi \,ff H$}

\begin{figure}
\begin{center}
\begin{tabular}{p{.45\linewidth}l}
\begin{minipage}{.25\linewidth}
    \includegraphics[scale=1.2]{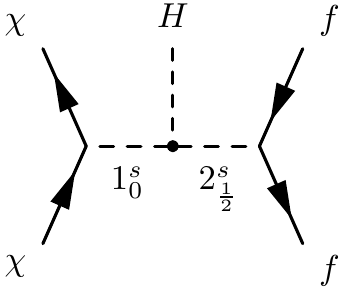}%
\end{minipage}
&
\begin{minipage}{.25\linewidth}
    \includegraphics[scale=1.2]{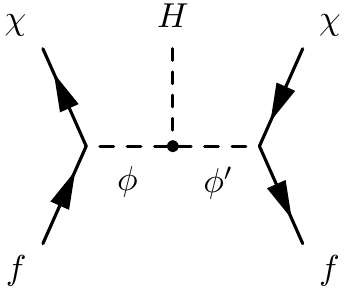}%
\end{minipage} \\ \ \\
\#B1 & \#B2 \\ \ \\ \ \\
\begin{minipage}{.25\linewidth}
    \includegraphics[scale=1.2]{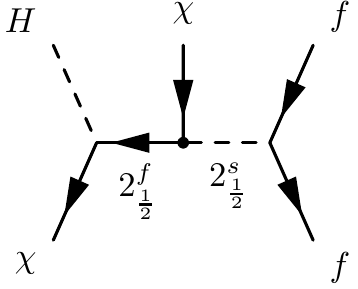}%
\end{minipage}
&
\begin{minipage}{.25\linewidth}
    \includegraphics[scale=1.2]{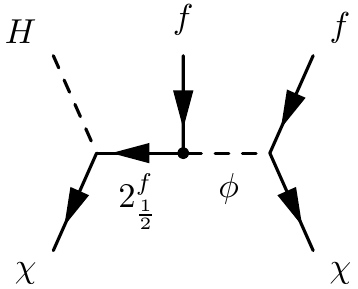}%
\end{minipage} \\ \ \\
\#B3 & \#B4 \\ \ \\ \ \\
\begin{minipage}{.25\linewidth}
    \includegraphics[scale=1.2]{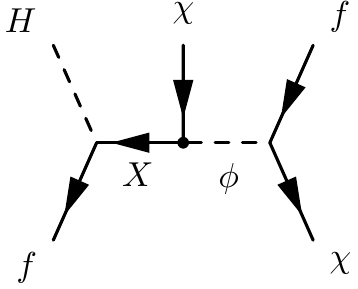}%
\end{minipage}
& \
\\ \ \\
\#B5 &  \ 
\end{tabular}
\end{center}
\caption{Different decompositions (Feynman diagrams) for the operator $\chi \chi ff H$. $\phi$ is a scalar and $X$ a fermion that has the quantum numbers of the external fermions $f$. The mediator $\phi^{(\prime)}$ corresponds to the mediators in \reffig{decompD6}.}
\label{fig_decompC}
\end{figure}

The decompositions of the operator $\chi \chi \,ff H$ are shown in \reffig{decompC}. Operators \#B2, \#B4 and \#B5 contain fields and couplings that again would also generate the operator $\chi \chi \,ff$ and are therefore not interesting to us. This is the same argument that has been made in the previous section. Operators \#B1 and \#B5 on the other hand contain a mediator $2^s_\frac{1}{2}$, which can be identified with the Higgs doublet, since it has the same quantum numbers. These operators are therefore actually implementations of the default Higgs-portal operators from \reffig{HPdecomp}. In conclusion, there exists no generic implementation of the $\chi \chi \,ff H$ that could be a \textit{dominant} contribution to DM interactions.

\subsection{Operators of the type $\chi \chi \,H^\dagger H S$}
\label{sec:cchhs}

In order to justify the suppression in \Equ\,(\ref{lambdahff}) we extend the SM with an extra scalar singlet $S$ 
and instead of the operator in \Equ\,(\ref{e3}) we will consider the following
\begin{equation}
\frac{1}{\Lambda^2} \chi\chi \,H^\dagger H\,S\,.\label{e5} 
\end{equation}

First we observe that in order to generate operators of the type $\chi \chi H^\dagger H S$ we need an additional scalar singlet which obtains a VEV. This would automatically imply the second operator from \reffig{HPdecomp}, where $S$ assumes the place of the mediator. We therefore require that the scalar $S$, as well as the DM field $\chi$, are charged under a discrete symmetry (in addition to the matter parity that stabilizes the DM particles) to forbid at least one of the necessary couplings. 
For example the field $S$ can have charge $\omega=\ee^{\ii\frac{2\pi}{3}}$ under a $\mathds{Z}_3$ symmetry. The DM then obtains a mass via the term $\lambda_S S\chi\chi$. In the case of Dirac DM this means $\chi_R^c \chi_L S$ is invariant for $q(\chi_L) = q(S) = \omega$ and $q(\chi_R)=\omega^2$ (so that $q(\overline{\chi_R}) = (\omega^2)^* = \omega$). In the case of Majorana DM we require $q(\chi)=\omega$. If $\chi$ is charged under the $\mathds{Z}_3$ symmetry with $q(\chi)=\omega$ we cannot have a Majorana mass term. When $S$ obtains a VEV, however, $\chi$ becomes effectively a Majorana fermion, due to the term $\lambda_S S\chi\chi$. We can easily check that now the operator $\chi \chi H^\dagger H$ ($\chi_R^c \chi_L H^\dagger H$ in the Dirac case) has charge $\omega^2$, while the operator $\chi \chi H^\dagger H S$ has charge $\omega^3 = 1$.

\begin{figure}[tp]
\begin{tabular}{cccc}
    \includegraphics[scale=1]{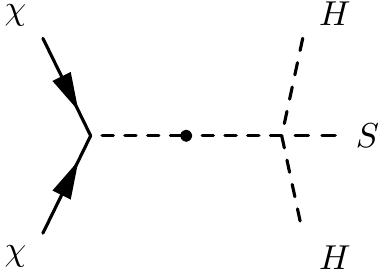}%
&
    \includegraphics[scale=1]{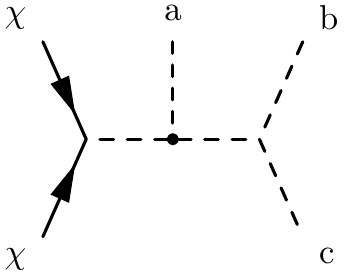}%
&
    \includegraphics[scale=1]{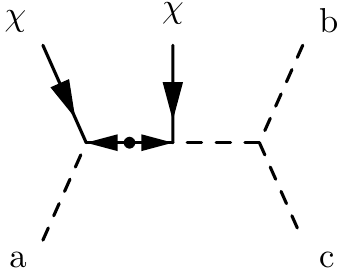}%
&
    \includegraphics[scale=1]{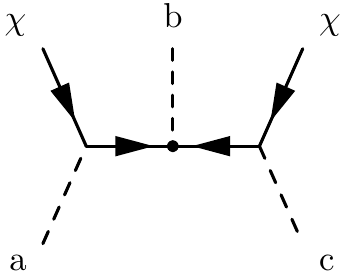}%
\\ \ \\
Topology c1 &
Topology c2 &
Topology c3 &
Topology c4
\end{tabular}
\caption{Different topologies for the decomposition of the effective operator $\chi \chi H^\dagger H S$.}
\label{fig_top}
\end{figure}

\begin{figure}[p]
\begin{center}
\begin{tabular}{p{.45\linewidth}l}
\begin{minipage}{.25\linewidth}
    \includegraphics[scale=1.2]{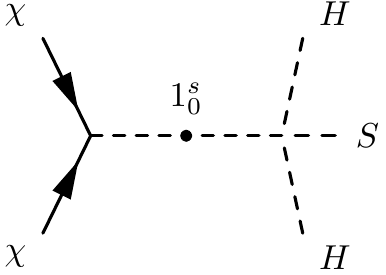}%
\end{minipage}
&
\begin{minipage}{.25\linewidth}
    \includegraphics[scale=1.2]{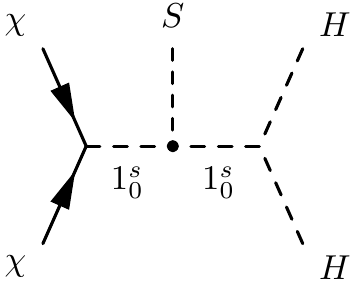}%
\end{minipage} \\ \ \\
\#C1 & \#C2 \\ \ \\
\begin{minipage}{.25\linewidth}
    \includegraphics[scale=1.2]{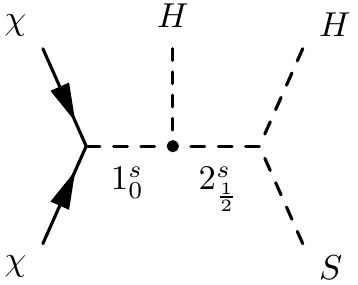}%
\end{minipage}
&
\begin{minipage}{.25\linewidth}
    \includegraphics[scale=1.2]{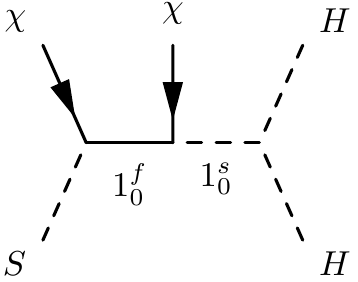}%
\end{minipage} \\ \ \\
\#C3 & \#C4 \\ \ \\
\begin{minipage}{.25\linewidth}
    \includegraphics[scale=1.2]{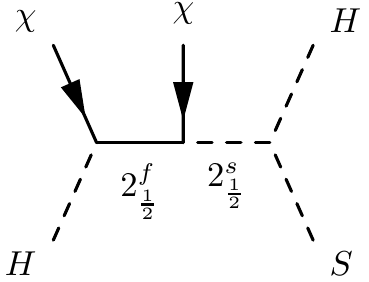}%
\end{minipage}
& 
\begin{minipage}{.25\linewidth}
    \includegraphics[scale=1.2]{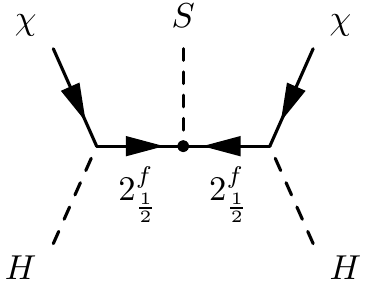}%
\end{minipage}\\ \ \\
\#C5 & \#C6 \\ \ \\
\begin{minipage}{.25\linewidth}
    \includegraphics[scale=1.2]{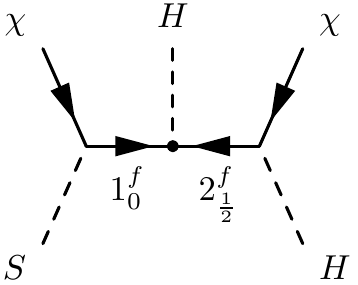}%
\end{minipage}
& \ \\ \ \\
\#C7 & \
\end{tabular}
\end{center}
\caption{Different decompositions (Feynman diagrams) for the operator $\chi \chi H^\dagger H S$.}
\label{fig_decomp}
\end{figure}

We will now discuss the possible decompositions of the operator $\chi \chi H^\dagger H S$. The possible topologies are shown in \reffig{top}. All decompositions are listed in \reftab{dXXHHS} and shown as Feynman diagrams in \reffig{decomp}.

\begin{table}[tp]
\begin{center}
\begin{tabular}{lcll}
\toprule
\ & Top. & ext. Fields & Mediators \\ \midrule
\#C1 & c1 & \ & $1^s_{0,+}$
\\
\#C2 & c2 & $a=S$, $b=H$, $c=H$ & $1^s_{0,+}$,  $1^s_{0,+}$ 
\\
\#C3 & c2 & $a=H$, $b=S$, $c=H$ & $1^s_{0,+}$,  $2^s_{\frac{1}{2},+}$
\\
\#C4 & c3 & $a=S$, $b=H$, $c=H$ & $1^f_{0,-}$, $1^s_{0,+}$
\\
\#C5 & c3 & $a=H$, $b=S$, $c=H$ & $2^f_{\frac{1}{2},-}$, $2^s_{\frac{1}{2},+}$
\\
\#C6 & c4 & $a=H$, $b=S$, $c=H$ & $2^f_{\frac{1}{2},-}$, $2^f_{-\frac{1}{2},-}$
 \\
\#C7 & c4 & $a=S$, $b=H$, $c=H$ & $1^f_{0,-}$, $2^f_{\frac{1}{2},-}$\\
\addlinespace[.2em]
\bottomrule
\end{tabular}
\caption{Decompositions of the operator $\chi\chi HHS$. The numbers in the first column correspond to the decompositions shown in \reffig{decomp}. The topologies (Top.) correspond to \reffig{top} where $a$, $b$ and $c$ are replaced accordingly. The last column lists the new mediator fields that have to be present in a model which generates this specific operator. The second subscript denotes the charge under the $\mathbb{Z}_2$ parity.}
\label{tab_dXXHHS}
\end{center}
\end{table}

As discussed earlier, we need to avoid the lower dimensional operators in \Figu~\ref{figintr}. As pointed out in section~\ref{sec_effOps}, it is not possible to forbid these operators by charging the external fields under a discrete symmetry, since some of these operators are invariant under any charge assignment. It is however possible to avoid these operators, if the mediator fields generating them are not present. So first of all we require that no additional vector fields are present, that couple to the DM particle. This is easy to achieve since we assume $\chi$ to be a gauge singlet. Furthermore we have to also avoid scalars with the quantum numbers specified in \refeq{qMediators}. One can easily check that none of the mediators in \reftab{dXXHHS} carries these quantum numbers. Note that the mediators $2^s_\frac{1}{2}$ of operators \#C3 and \#C5 differ from the mediators generating the operator $\mathcal{O}_1$ only due to the $\mathbb{Z}_2$ symmetry. 

In summary, we need the following ingredients: (i) A $\mathbb{Z}_2$ symmetry that stabilizes the DM, (ii) an additional $\mathbb{Z}_3$ in order to forbid the leading order Higgs-portal interaction $\chi\chi H^\dagger H$ and (iii) we need to avoid certain mediators to forbid the leading order direct interaction $\chi\chi ff$. 
An interesting observation can be made here regarding point (iii). It is possible to suppress the interactions of the DM with only specific SM fields. We could, for example, only forbid colored mediators but allow for SU(3) singlet mediators that generate operator $O_1$. As a consequence, we would have interactions between DM and Leptons via the operator $\chi\chi ff$, whereas interactions to quarks, which are relevant for direct detection, would be suppressed, since they are only generated by the Higgs-portal process via $\chi\chi H^\dagger H S$. This would give us a model where DM is leptophilic, meaning that it preferably interacts with leptons and not with quarks. Such models have been discussed in literature recently, see,\eg, Refs.~\cite{Haba:2010ag,Lopez-Honorez:2013wla}.

There are some remarks regarding the different decompositions. We have the coupling $\lambda_\chi \, \chi \chi S$. In a non-SUSY scenario we cannot forbid the terms $m_S S^\dagger S$ and $\lambda_H S^\dagger S H^\dagger H$. This will automatically generate op. \#C1 with
\begin{align}
  \lambda_\chi \lambda_H \frac{v_S}{m_S^2} H^\dagger H \chi \chi = \lambda_H \frac{m_\chi}{m_S^2} H^\dagger H \chi \chi\,,
\end{align}
where $1^s_{0,\omega}=S$. So this decomposition will be present (possibly besides others) in any model that generates the operator $\chi \chi HHS$. 

Looking at decomposition \#C4, we face another possible problem: the mediator $1^f_{0,\omega^2}\equiv \chi'$ and the coupling $S \chi \chi'$ can introduce a Dirac mass-term  $m_\chi' \chi \chi'$. This means we would obtain another DM $\chi'$ component which mixes with $\chi$ and allow for the leading order operator $\chi \chi' H^\dagger H$. This decomposition is therefore problematic.

\subsection{Operators of the type $\chi \chi \,(H^\dagger H)^2$}

\begin{figure}[p]
\begin{center}
\begin{tabular}{p{.45\linewidth}c}
    \includegraphics[scale=1.1]{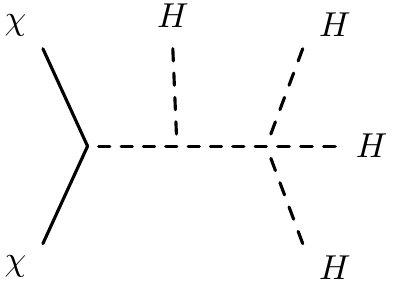}%
&
    \includegraphics[scale=1.1]{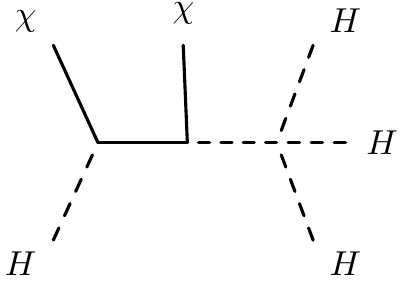}%
\\ \ \\
\hspace*{.7cm} Topology e1a & Topology e1b
\\ \ \\ \ \\
    \includegraphics[scale=1.1]{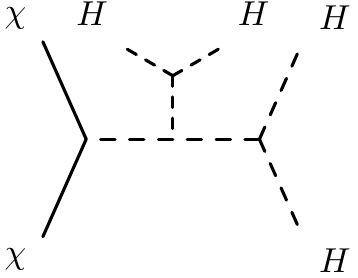}%
&
    \includegraphics[scale=1.1]{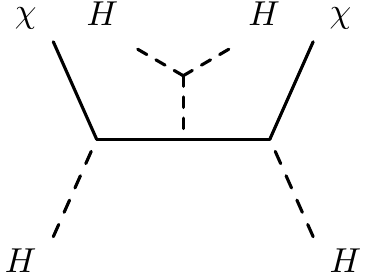}%
\\ \ \\
\hspace*{.7cm} Topology e2a & Topology e2b
\\ \ \\ \ \\
\end{tabular}  
\begin{tabular}{p{.35\linewidth}p{.35\linewidth}c}
    \includegraphics[scale=1.1]{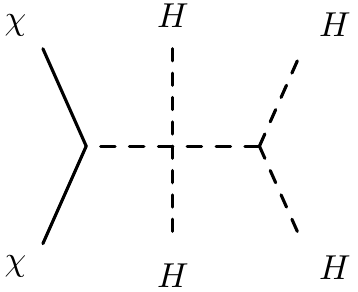}%
&
    \includegraphics[scale=1.1]{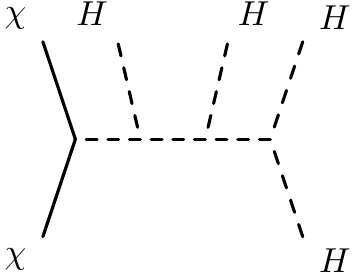}%
&
    \includegraphics[scale=1.1]{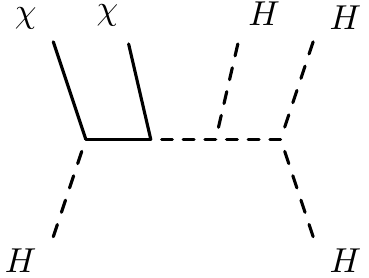}%
\\ \ \\
\hspace*{.5cm} Topology e3  & \hspace*{.6cm} Topology e4a & Topology e4b
\\ \ \\ \ \\
    \includegraphics[scale=1.1]{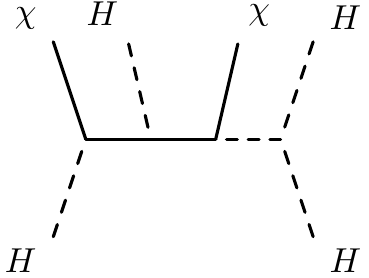}%
&
   \includegraphics[scale=1.1]{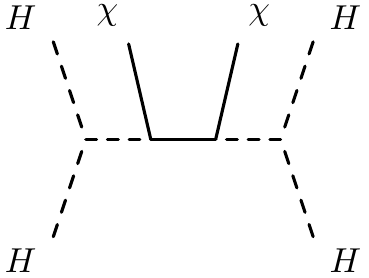}%
&
   \includegraphics[scale=1.1]{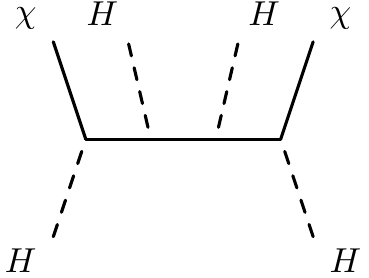}%
\\ \ \\
\hspace*{.5cm} Topology e4c & \hspace*{.6cm} Topology e4d & Topology e4e
\end{tabular}
\caption{Different topologies for the decompositions of the effective operator $\chi \chi\,(H^\dagger H)^2$.}
\label{fig_topH4}
\end{center}
\end{figure}

We now want to discuss the $d=7$ operator $\chi \chi \,(H^\dagger H)^2$. This is the next-to leading operator in the Higgs-portal scenario that contains only SM fields. The possible topologies for the operator $\chi\chi (H^\dagger H)^2$ are presented in \reffig{topH4}. As we will show in the following, all decompositions of this operator are difficult to obtain as a dominant contribution to Higgs-portal induced DM interactions.

This is due to various reasons. In case of topologies e2, e3 and e4a-e4d we always have a vertex where two external Higgs fields couple to a mediator, which can be a SU(2) singlet or triplet. In the singlet case we encounter the problem we now from the discussion of \reffig{HHvS}: If the singlet mediator will obtain a VEV we actually have an operator of the type $\chi \chi H^\dagger H S$ and not $\chi\chi (H^\dagger H)^2$. In fact, we cannot avoid this VEV due to the coupling of the singlet to the Higgs, see also Refs.~\cite{Krauss:2011ur,Krauss:2013gy} for a similar discussion for higher dimensional operators generating neutrino mass. In the case of a triplet mediator that couples to a pair of external Higgs fields we can argue analogously: The neutral component of the triplet also obtains a VEV via its Higgs coupling and again we have effectively a lower dimensional operator. In the triplet case, we know however, that the VEV and the coupling must be small, since otherwise we would generate too large neutrino masses via a type-II seesaw diagram.

Also topology e1 has a similar problem: The quartic coupling of the three external Higgs fields to a scalar mediator implies that this mediator is indeed also a Higgs field, since it is an SU(2) doublet with the according quantum numbers. Hence we can cut this diagram at this mediator and obtain the standard Higgs-portal operator $\mathcal{O}_H$.

The only remaining option is topology e4e. Since it has only fermionic mediators we do not run into the same difficulties as discussed above. Here, however, different ones appear: We have the the mediator coupling to the Higgs and and the $\chi$ field is the same that generates the operator $\#H1$ in \reffig{HPdecomp}. So we cannot avoid the standard Higgs-portal operator. Even an additional symmetry would be of no help, since $H^\dagger H$ is a singlet under any such symmetry.

In summary, the operator $\chi \chi \,(H^\dagger H)^2$ is not a good option for generating DM interactions at higher dimensions.

\section{LHC phenomenology}
\label{sec:lhc}

\begin{figure}
 \begin{center}
 \includegraphics[scale=1.1]{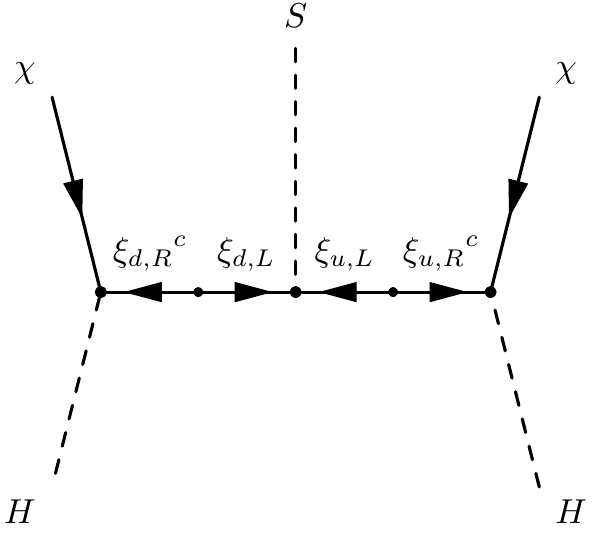}%
\caption{Decompositions of the effective $d=6$ operator \#C6.}
\label{fig:decomp01}
 \end{center}
\end{figure}

In this section we exemplify generic LHC signatures for such models using one specific example
(operator \#C6), which is shown in in \Figu~\ref{fig:decomp01}, with the fields listed in \Tab~\ref{tab:charges01}.
The new fermions all belong to the  $2^f_{\pm\frac{1}{2}}$ representation, except from the DM candidate $\chi$ which is an SM singlet.
\begin{table}
 \begin{center}
 \begin{tabular}{lccccccc}
\toprule
   Fields:	& SM 	&\multicolumn{5}{c}{Fermions} & Scalars\\ \cmidrule(rl){3-7}
		&  	& $\xi_{d,L}$ & $\xi_{u,L}$ & $(\xi_{u,R})^c$ 	& $(\xi_{d,R})^c$ 	& $\chi$ 	& $S$ \\ \midrule
   $SU(2)$	& 	& 2	& 2	&  2 & 2	& 1		& 1 \\
   $U(1)_Y$	& 	& $-\frac{1}{2}$ & $+\frac{1}{2}$ & $-\frac{1}{2}$ & $+\frac{1}{2}$ & 0 & 0\\ 
  $\mathbb{Z}_3$& 1 	& $\omega$ &  $\omega$ &  $\omega^2$ &  $\omega^2$ & $\omega$ &  $\omega$ \\ 
  $\mathbb{Z}_2$& + 	& - 	& -	& - 		&  - 		& - 		&  + \\ 
\bottomrule
  \end{tabular}
 \end{center}
\caption{Field charges, with $\omega = \ee^{\ii \frac{\pi}{3}}$ under $\mathbb{Z}_3$, whereas ``+'' signifies a phase of 1 and ``-'' signifies a phase of $\ee^{\ii \pi}$ under the $\mathbb{Z}_2$.}
\label{tab:charges01}
\end{table}
The according Lagrangian reads:
\begin{align}
\begin{split}
 \mathcal{L}_\text{\#C6} = \, &\mathcal{L}_\text{SM} + \left[\lambda_{\xi_{u}}\, H \cdot (\xi_{u,R})^c  \,\chi
+ \lambda_{\xi_d}\, H^\dagger \cdot (\xi_{d,R})^c \, \chi
+ \lambda_{\xi \xi S}\, S\, \xi_{u,L} \, \cdot\, \xi_{d,L}\right . \\ &
+ \lambda_{\xi \xi S}'\, S^*\, (\xi_{u,R})^c \, \cdot \, (\xi_{d,R})^c 
+ m_u\,  \xi_{u,L} \cdot  (\xi_{u,R})^c
+ m_d\, (\xi_{d,R})^c \cdot \xi_{d,L} \\& \left. 
+ \lambda_{S\chi\chi} S \chi \chi +  \kappa_{S} S^3 + \hc \right] \\&
+ \lambda_{SSHH} (S^* S) (H^\dagger H)
+ m_S S^* S + \lambda_{S} (S^* S)^2\,.
\end{split}
\end{align}
The ``$\cdot$'' denotes the invariant product of the SU(2) doublets. The doublets can explicitly be expressed as
\begin{align}
\xi_{d,L} &= \vectwo{\xi^0_{d,L}}{\xi^-_{d,L}}  & \xi_{u,L}&=\vectwo{\xi^+_{u,L}}{\xi^0_{u,L}} & 
\xi_{u,R}{}^c &= \vectwo{\xi^0_{u,R}{}^c}{(\xi_{u,R}{}^c)^-} & \xi_{d,R}{}^c &= \vectwo{(\xi_{d,R}{}^c)^+}{\xi^0_{d,R}{}^c} \,.
\end{align}
Several aspects can already be understood from the mass
spectrum of the additional particles. Here we will assume for simplicity
that all new couplings are real.
In addition to the SM-fermions this model contains two charged fermions and five neutral Majorana fermions.  We get
the mass matrix
\begin{eqnarray}
M_{X^+} = \left( \begin{array}{cc}
m_{\xi_u} & -\lambda'_{\xi \xi S} \frac{v_S}{\sqrt{2}}\\
\lambda_{\xi \xi S} \frac{v_S}{\sqrt{2}}  & m_{\xi_d}
\end{array} \right)
\end{eqnarray}
 for the charged ones in the basis $(\xi_u,\xi_d^c)$ and 
\begin{eqnarray}
M_{X^0} = \left( \begin{array}{ccccc}
 0 & -m_{\xi_u} &  0& -\lambda_{\xi \xi S}  \frac{v_S}{\sqrt{2}} & 0\\
-m_{\xi_u} & 0 &  \lambda_{\xi \xi S}'  \frac{v_S}{\sqrt{2}} &0&  -\lambda_{\xi u}  \frac{v}{\sqrt{2}} \\
0 & \lambda_{\xi \xi S}'  \frac{v_S}{\sqrt{2}} & 0 &  -m_{\xi_d} & \lambda_{\xi d}  \frac{v}{\sqrt{2}} \\
 -\lambda_{\xi \xi S}  \frac{v_S}{\sqrt{2}} & 0 &  -m_{\xi_d} & 0 & 0 \\
0 &   -\lambda_{\xi u}  \frac{v}{\sqrt{2}} &  \lambda_{\xi d}  \frac{v}{\sqrt{2}} & 0 & 
\lambda_{S\chi\chi} \frac{v_S}{\sqrt{2}} 
\end{array} \right)
\end{eqnarray} 
for the neutral ones in the basis 
$(\xi^0_{u,L}, \xi^0_{u,R}{}^c, \xi^0_{d,R}{}^c, \xi^0_{d,L},\chi)$.
Here $\langle H^0 \rangle = v/\sqrt{2}$ and $\langle S \rangle = v_s/\sqrt{2}$. 
We denote the mass eigenstates of the charged fermions by $X^+_i$ ($i=1,2$) and the neutral ones by $X^0_j$ 
$(j=1,\dots,5)$.

In the scalar sector one has
a massive pseudoscalar $P^0$, which is purely singlet-like, with mass $m^2_P = -\frac{9 \kappa v_S}{\sqrt{2}}$.
Note, that this implies a relative sign between $v_S$ and $\kappa$.
Moreover, there are two scalar Higgs bosons with a mass matrix
\begin{equation}
M^2_H = \left( \begin{array}{cc}
 2 \lambda v^2 & \lambda_{SSHH} v v_S \\
 \lambda_{SSHH} v v_S & 2 \lambda_S v_S^2+\frac{3 \kappa v_S}{\sqrt{2}}
\end{array} \right)
=  \left( \begin{array}{cc}
 2 \lambda v^2 & \lambda_{SSHH} v v_S \\
 \lambda_{SSHH} v v_S & 2 \lambda_S v_S^2 - \frac{1}{3} m^2_P
\end{array} \right)
\end{equation}
Note, that this gives an upper bound on $m^2_P \le (6 \lambda_S - \frac{3}{2\lambda} \lambda_{SSHH}^2) v_S^2$ at tree level as the determinant has to be positive. We denote the mass eigenstates by $h_i$ ($i=1,2$). One of them has to be essentially the
SM Higgs bosons as required by existing LHC data \cite{ATLAS:2013sla,CMS:2013wda}
and as a consequence the second one will be mainly a singlet like states which can hardly be produced directly.
However, as we will see below, it can be produced in the cascade decays of the additional fermions.
Depending on the mass hierarchy the pseudoscalar $P$ can decay either into two of the new fermions or into two photons via a loop of $\xi^+$ fields. For the latter the relevant couplings are $\lambda_{\xi \xi S}$ and $\lambda_{\xi \xi S}'$ respectively. 
In case that the photon decay mode is dominating, one has to check that its life-time is short enough in order not to destroy the successful prediction of Big Bang Nucleosynthesis. The corresponding partial decay width is approximately
\begin{align}
 \Gamma(P\rightarrow\gamma\gamma) \approx (\lambda_{\xi \xi S}+\lambda_{\xi \xi S}')^2 \times \mathcal{O}(\text{keV})\,,
\end{align}
for $m_{X^+}=500$ GeV.

For the LHC phenomenology one can distinguish two limiting cases:
$m_{\xi_d,\xi_u} \gg \lambda_{\xi \xi S}  \frac{v_S}{\sqrt{2}}, \lambda'_{\xi \xi S}  \frac{v_S}{\sqrt{2}}$ (case A) and 
$m_{\xi_d,\xi_u} \ll \lambda_{\xi \xi S}  \frac{v_S}{\sqrt{2}}, \lambda'_{\xi \xi S}  \frac{v_S}{\sqrt{2}}$ (case B). 
Note that the case
$m_{\xi_d,\xi_u} \simeq \lambda_{\xi \xi S}  \frac{v_S}{\sqrt{2}} \simeq \lambda'_{\xi \xi S}  \frac{v_S}{\sqrt{2}}$ 
is heavily constrained by the 
LEP and Tevatron data:  in this case one of the charged fermions would be rather light which is excluded
by the existing data \cite{Beringer:1900zz}. As the lightest neutral fermion should be mainly the singlet
state one gets two conditions on the underlying couplings:  (i) $\lambda_{S\chi\chi}$
is sufficiently small such that the lightest neutral fermion is essentially a singlet and (ii) 
 $\lambda_{\xi u}$ and $\lambda_{\xi d}$ are sufficiently small compared to the other couplings
such that the mixing of the singlet-like state  with
the neutral components of the $SU(2)$ doublets remains  small. 

In case A the charged fermions will have masses close to $m_{\xi_d}$ and $m_{\xi_u}$ and also 
the neutral sector will contain two pseudo-Dirac fermions with masses close to these values. 
The parameters $\lambda_{\xi \xi S}$ and $\lambda_{\xi \xi S}'$ are bounded from below, since they must guarantee small enough life-times of the pseudoscalar, as discussed above.
Now if $m_{\xi_d} \simeq  m_{\xi_u}$ then the dominant decay modes are
\begin{eqnarray}
\label{eq:chip_decays_A}
X^+_i &\to& W^+ X^0_1 \,\, (i=1,2) \\
X^0_j &\to& Z X^0_1 \,\,,\,\, h_i  X^0_1 \,\,,\,\, P  X^0_1 \,\, (j=2,3,4,5)
\label{eq:chiz_decays_A}
\end{eqnarray}
where $X^0_1 \simeq \chi$. Potentially this will lead to displaced vertices if 
$\lambda_{\xi u}$ and $\lambda_{\xi d}$ are sufficiently small. Note, that these decays
will be the dominant source for the additional Higgs bosons. In case that either
$m_{\xi_d}$ is much larger than $m_{\xi_u}$ of vice versa, then the following decay modes
are open in addition to the above ones:
\begin{eqnarray}
X^+_2 &\to& Z X^+_1 \,\,,\,\, h_i  X^+_1 \,\,,\,\, P  X^+_1 \,\,,\,\, W^+ X^0_j \,\,(j=2,3) \\
X^0_j &\to& Z X^0_k \,\,,\,\, h_i  X^0_k \,\,,\,\, P  X^0_k \,\,,\,\, W^\pm X^\mp_1 \text{ with }
j=4,5 \text{ and } k=2,3
\end{eqnarray}
In both cases this looks similar to the production and decays of charginos and neutralinos
in the context of supersymmetric models. At the LHC the new
fermions which are essentially members of the $SU(2)$ doublets are produced via
Drell-Yan processes and can in principle be detected with masses of up to 
about 800 GeV, see,\eg, Ref.~\cite{Krauss:2011ur}.
For completeness we note, that these decays will occur via
off-shell vector bosons and off-shell Higgs bosons
if the mass gaps between the new fermions are so small that all two-body decays are kinematically suppressed.

In case B one finds that there are two charged fermions and two neutral quasi-Dirac fermions
with masses close to $\lambda_{\xi \xi S}  \frac{v_S}{\sqrt{2}}$. Since the main contribution to $m_\xi$ arises from $\lambda_{\xi \xi S}$ and $\lambda'_{\xi \xi S}$, the lower limit on $m_\xi$ guarantees a large coupling, which implies a short life-time of the pseudoscalar. The corresponding decays
are the same as above in  case A with $m_{\xi_d} \simeq  m_{\xi_u}$, see \Equs~(\ref{eq:chip_decays_A})
and (\ref{eq:chiz_decays_A}).

\section{Summary and conclusions}

We have demonstrated that models exist where the dominant contribution to interactions of DM particles with ordinary matter is generated by a higher dimensional operator. In this study  we have systematically decomposed the leading higher dimensional operators that induce direct DM detection. We have also shown that Higgs-portal interactions have to be taken into account as well, since they may generate direct DM detection at a similar order. 

We have specifically discussed fermionic DM models where the lowest dimensional effective operators for interactions between the DM particle and SM particles, $\chi \chi\, ff$ and $\chi \chi H^\dagger H$, are not present. Various possible UV completions of the higher dimensional operators imply the existence of a scalar singlet $S$. Therefore we have discussed the next-to leading order direct interaction operators $\chi \chi\, ff S$ and $\chi\chi ff H$ as well as the next-to leading order Higgs-portal operators $\chi \chi H^\dagger H S$ and $\chi \chi (H^\dagger H)^2$. 
The first two of these cannot be a dominant contribution to DM interactions, because they would induce effective interactions of a lower dimension. 
A similar argument has been presented for the higher dimensional effective operator $\chi \chi (H^\dagger H)^2$. 
Hence we concluded that the lowest order generic possibility requires in fact Higgs-portal interactions induced by an effective $d=6$ operator $\chi \chi H^\dagger H S$. This can be achieved in scenarios with a $\mathbb{Z}_2$ symmetry that stabilizes the DM particle and an additonal $\mathbb{Z}_3$ symmetry to forbid the leading Higgs portal operator. The cross-sections of processes induced by this operator are suppressed as compared to the lowest dimensional Higgs-portal interaction. Such a scenario is a possible reason, why no conclusive evidence of DM has been found so far. We have also shown that interactions with leptons can still be generated by less suppressed operators if color singlets are chosen as mediators. Such leptophilic DM can guarantee DM relic abundances that are in accordance with cosmological bounds.

In a particularly interesting realization of the operator $\chi \chi H^\dagger H S$ all mediators are fermions. These new particles can be produced via Drell-Yan processes at the LHC and have decays similar to those of charginos and neutralinos. 
By comparison with similar models we expect that these particles can be discovered at the LHC for masses up to 800 GeV. This will be the content of a future study.

\section*{Acknowledgments}

This work has been supported  by  DFG projects GRK-1147, PO-1337/2-1, WI-2639/3-1, and WI-2639/4-1.
This work has also been supported by the FP7 Invisibles network ``Neutrinos, dark matter, and dark energy physics'' (Marie Curie
Actions, PITN-GA-2011-289442), the ``Helmholtz Alliance for Astroparticle Physics HAP'', funded by the Initiative and Networking fund of the Helmholtz association.

\newpage

%

\end{document}